# Instrumentation for ESO's Extremely Large Telescope

Suzanne Ramsay[1]
Michele Cirasuolo[1]
Paola Amico[1]
Nagaraja Naidu Bezawada[1]
Patrick Caillier[1]
Frédéric Derie[1]
Reinhold Dorn[1]
Sebastian Egner[1]
Elizabeth George[1]
Frédéric Gonté[1]
Peter Hammersley[1]
Christoph Haupt[1]
Derek Ives[1]
Gerd Jakob[1]
Florian Kerber[1]
Vincenzo Mainieri[1]
Antonio Manescau[1]
Sylvain Oberti[1]
Celine Peroux[1]
Oliver Pfuhl[1]
Ulf Seemann[1]
Ralf Siebenmorgen[1]
Christian Schmid[1]
Joël Vernet[1]
and the ESO ELT Programme
  and follow-up team

[1] ESO

Design and construction of the instruments for ESO's Extremely Large Telescope (ELT) began in 2015. We present here a brief overview of the status of the ELT Instrumentation Plan. Dedicated articles on each instrument are presented elsewhere in this volume.

## Instruments planned for ESO's ELT

When, in December 2014, the ESO Council gave the green light for the construction of the 39-m Extremely Large Telescope[1] in two phases (de Zeeuw, Tamai & Liske, 2014), this triggered the final preparations to launch the design and construction of the powerful instrument suite for this telescope. The ELT Instrumentation Plan, to provide the instruments to meet the science case for the telescope, had already been defined in consultation with ESO's science community and scientific and technical advisory committees. The instruments were selected following a set of Phase A conceptual design studies that have been described previously (see papers in The Messenger 140, 2010). While ESO is ultimately responsible for delivering the instruments to the scientific community on time and with the expected performance, an important feature of the Instrumentation Plan is that the instruments are being developed in collaboration between ESO and consortia made up of universities and institutes in the Member States and beyond. This model has worked very successfully for the delivery of instruments to the Very Large Telescope (VLT) and is a key aspect of the interaction between ESO as an organisation and the astronomical community. Figure 1 shows a timeline for instrument development as it stands at the time of writing.

A pair of instruments was selected by the ELT Science Working Group to be delivered for first light: the High Angular Resolution Monolithic Optical and Near-infrared Integral field spectrograph (HARMONI) and the Multi-adaptive optics Imaging CamerA for Deep Observations (MICADO), a near-infrared camera. Adaptive optics systems tailored to meet the scientific goals of each of these instruments are also being developed. The HARMONI Consortium is building a laser tomographic adaptive optics (LTAO) module. A multi-conjugate adaptive optics (MCAO) module, the Multi-conjugate Adaptive Optics RelaY (MAORY), is being developed as a facility adaptive optics system with two "clients" — MICADO and a future multi-object spectrograph. Together this first light pair of workhorse instruments will immediately exploit both the enormous collecting area and the superb spatial resolution of the new telescope, enabling a wide range of scientific projects to be executed at first light. The next instrument in the Instrumentation Plan is the Mid-infrared ELT Imager and Spectrograph (METIS), working in the mid-infrared (3–14 μm) with single-conjugate adaptive optics (SCAO). All of these instruments are formally part of ESO's ELT Construction Programme. Agreements for the design, construction and commissioning of the three instruments plus MAORY were signed in 2015. The LTAO module for HARMONI was one of the Phase 2 items whose funding was initially deferred (de Zeeuw, Tamai & Liske, 2014) and so only the work to carry out the preliminary design was included in the agreement for HARMONI. Formal approval of the next design phases and the construction of the LTAO module was signed in 2019 when funds for this module became available. It will now be delivered along with HARMONI for first light with the spectrograph, ensuring optimised performance and increased sky coverage.

The instruments under construction have now completed the important preliminary design phase, during which the basic concept for the instrument is refined and compliance with the scientific and technical requirements is confirmed. The first Preliminary Design Review (PDR) meeting, for HARMONI, was held in November 2017, those for MICADO and METIS followed in October 2018 and May 2019, respectively. Everything about these projects is on a very large scale, as befits the extreme size of the telescope. The effort that goes into the PDRs is no exception. The document package for each instrument amounts to more than one hundred documents and many thousands of pages. The design concepts have been reviewed by tens of engineers from ESO with support from external experts from industry and from other extremely large telescope projects, such as the Thirty Meter Telescope (TMT)[2] and the Giant Magellan Telescope (GMT)[3]. Each of these instruments is now formally in the final design phase during which the design is detailed to the level that manufacturing of the key components can start after the Final Design Review (FDR) is concluded. The design for MAORY has undergone significant revision since the Phase A study; it has been optimised for manufacturability and ease of alignment, compliance with the available volume and mass, and also to ensure that it provides a good interface for the two client instruments. The PDR for MAORY is planned for the second quarter of 2021.

As the instrument designs have progressed, much has been learnt about the real resource requirements of these huge systems with their challenging performance specifications. Mass and power budgets, space envelopes, vibration control and maintenance requirements are major topics of discussion. Careful follow-up and management of these items has allowed MICADO, HARMONI and METIS to move into their FDR phases without any loss of functionality or performance,

| Instrument | Main specifications | | | Schedule | | | | |
|---|---|---|---|---|---|---|---|---|
| | Field of view/slit length/ pixel scale | Spectral resolution | Wavelength coverage (µm) | Phase A | Project start | PDR | FDR | First light |
| MICADO | Imager (with coronagraph) 50.5″ × 50.5″ at 4 mas/pix 19′ × 19′ at 1.5 mas/pix | I, Z, Y, J, H, K + narrowbands | 0.8–2.45 | 2010 | 2015 | 2019 | | |
| | Single slit | R ~ 20 000 | | | | | | |
| MAORY | AO Module SCAO – MCAO | | 0.8–2.45 | 2010 | 2015 | | | |
| HARMONI + LTAO | IFU 4 spaxel scales from: 0.8″ × 0.6″ at 4 mas/pix to 6.1″ × 9.1″ at 30 × 60 mas/pix (with coronagraph) | R ~ 3200 R ~ 7100 R ~ 17 000 | 0.47–2.45 | 2010 | 2015 | 2018 | | |
| METIS | Imager (with coronagraph) 10.5″ × 10.5″ at 5 mas/pix in L, M 13.5″ × 13.5″ at 7 mas/pix in N | L, M, N + narrowbands | 3–13 | 2010 | 2015 | 2019 | | |
| | Single slit | R ~ 1400 in L R ~ 1900 in M R ~ 400 in N | | | | | | |
| | IFU 0.6″ × 0.9″ at 8 mas/pix (with coronagraph) | L, M bands R ~ 100 000 | | | | | | |
| HIRES | Single object | R ~ 100 000 | 0.4–1.8 simultaneously | 2018 | | | | |
| | IFU (SCAO) | | | | | | | |
| | Multi object (TBC) | R ~ 10 000 | | | | | | |
| MOSAIC | ~ 7-arcminute FoV ~ 200 objects (TBC) | R ~ 5000–20 000 | 0.45–1.8 (TBC) | 2018 | | | | |
| | ~ 8 IFUs (TBC) | R ~ 5000–20 000 | 0.8–1.8 (TBC) | | | | | |
| PCS | Extreme AO camera and spectrograph | TBC | TBC | | | | | |

1 milliarcsecond (mas) = 0.001″

Figure 1. The ELT Instrumentation roadmap and timeline.

despite some greatly increased demands on the telescope and the observatory. MAORY is also on track to meet its requirements as the PDR approaches. The lessons learnt from these pioneering instruments are being applied to the development of future instruments.

In addition to the first three instruments and their adaptive optics modules, the ELT Construction Programme included two Phase A studies, for a multi-object spectrograph (named MOSAIC), and a high spectral resolving power, high-stability spectrograph (named HIRES). The original Phase A design studies carried out from 2007 to 2010 included three separate concepts for a multi-object spectrograph (OPTIMOS-EVE, Hammer, Kaper & Dalton, 2010; OPTIMOS-DIORAMAS, Le Fèvre et al., 2010; and EAGLE, Morris & Cuby, 2010) and two for a high resolving power spectrograph (CODEX, Pasquini et al., 2010 and SIMPLE, Origlia, Oliva & Maiolino, 2010). In 2016 ESO issued a call for two Phase A studies for HIRES and MOSAIC in order to update and optimise the scientific scope and specifications of these instruments, taking into account how best to complement the observing capabilities offered by the first-light instruments. These instrument studies concluded in 2018.

The next stage of construction of HIRES and MOSAIC, and the funding of the future ELT Planetary Camera and Spectrograph (ELT-PCS), fall outside the ELT Construction Programme and within the Armazones Instrumentation Programme (AIP). The AIP will manage all future instrument development during the lifetime of the ELT. The agreements for the construction phase of MOSAIC and HIRES, including the detailed scientific requirements, are being finalised now. ESO's committees support the start of the construction of these instruments once the resources (funding, effort and Guaranteed Time) needed to complete the first instruments are well understood and secured. This milestone is expected when the last of the PDRs for the first instruments is complete. An important step towards the launch of the MOSAIC and HIRES construction phases was the recent approval by the ESO Council for the procurement of the second prefocal station for the Nasmyth B platform that will host MOSAIC and HIRES. Taken together, the instruments so far planned for the ELT offer excellent coverage of the observing parameter space, allowing astronomers to tackle a very broad range of science cases that will fully exploit the collecting power and diffraction limit of the ELT. As shown in Figure 2, users will have access to imaging and spectroscopy, across a wide range of wavelengths and spectral resolving powers, in a variety of observing modes, and including high-contrast, precision astrometry and non-sidereal tracking.

ELT-PCS is the planet hunter that will deliver one of the highest priority and most challenging science goals of the telescope — the detection and characterisation of exo-Earths. Given the rapidly changing understanding of the population of exoplanets and the many new facilities that are being developed to study them, it was decided in 2010 that ELT-PCS should start later in the overall timeline in order to allow for developments in the science case. Furthermore, achieving the extreme contrast ratios required for these observations requires research and development in the field of adaptive optics and coronagraphy. Prototyping of components that are needed for ELT-PCS is part of ESO's ongoing Technology Development programme.

The development of this instrument is linked to both the level of technical readiness of these prototypes and the availability of funding and effort.

In other articles in this issue details of the science case, operational modes and instrument concepts are given for each of the instruments.

## Activities at ESO

The activities at ESO that support the development of the instruments for the ELT take a number of different forms. To ensure that ESO meets its commitments for the delivery of the instruments, a dedicated follow-up team of scientists, managers and engineers across all disciplines is assigned to work with each instrument team. The role of this follow-up team is to support the consortia with their expertise and also with understanding the interface to and performance of the telescope. This team also provides each instrument consortium with guidance on the application of the ESO standards. Standardisation of hardware and software across the observatories is crucial for cost- and time-effective operation and maintenance of the telescope(s) and instruments and is a significant development activity for ESO. The ELT standards include cryogenic components, control and dataflow software, instrument control electronics, real-time computing and wavefront sensor cameras. The standards have been either adopted or extended from the Paranal Observatory standards, or are new developments that may also be adopted by new instruments for Paranal when that is technically feasible.

Engineers and scientists also work within the consortia to deliver specific components or expertise and so ESO is also an associate member of each instrument consortium. ESO has world-leading expertise in detector technology and traditionally delivers the science detectors with standard detector controllers to the instruments on the VLT, and the same concept has been adopted for the ELT instruments. For its optical mode, HARMONI will use the Teledyne-e2V CCD231-84 deep-depletion silicon CCDs already used in the Multi Unit Spectroscopic Explorer (MUSE). Both MICADO and HARMONI will use the Hawaii 4RG detector from Teledyne-e2V for their near-infrared modules. METIS will use near-infrared detectors from the Hawaii "family", the Hawaii 2RG, for its *LM*-band imager and spectrometer. A particularly exciting development for METIS is that it will use a new detector for the *N*-band observations. The initial plan was to use the Aquarius detector that has been used on-sky with the VLT Imager and Spectrometer for mid-InfraRed (VISIR). However, the technology of the new GeoSNAP detector from Teledyne-E2V is now sufficiently ready that the decision to switch to this detector was taken after the METIS PDR. Simplifications to the instrument design come from this change but, most importantly, the observing efficiency in the *N*-band imaging mode, where many of the important science cases in exoplanets will be tackled, is expected to be many orders of magnitude higher than with the design using the Aquarius detector. ESO leads the work package for the GeoSNAP detector that will be tested at the METIS consortium partners the Max Planck Institute for Astronomy and the University of Michigan. Finally, an update of the standard detector controller, the Next Generation Controller (NGC), to a new edition (NGCII) with enhanced performance and matching the interface requirements of the new telescope is being carried out under the Technology Development Programme.

Expertise in adaptive optics is also an important input to the instrument consortia. In this regard, ESO engineers and physicists work within the instrument consortia, fully integrated into the teams, providing backup for simulating the telescope behaviour and instrument performance, developing the calibration strategies for the adaptive optics and contributing to the engineering design of the adaptive optics modules based on their knowledge of the ELT and experience from the Adaptive Optics Facility upgrade programme. ESO is also leading an effort to coordinate the expertise of all the groups working on SCAO for the ELT, including for the telescope, to explore common solutions for the calibration of these systems.

ESO maintains an overview of all of the systems on the telescope to ensure a fully working system and is responsible for the interface from all instruments to the observatory and between MAORY and its client instruments. One of the challenges facing both the instrument consortia and ESO is the parallel development of the telescope and the instrumentation. The agreements that have been signed with the instrument consortia include formal documentation describing the interface to the telescope systems and the requirements for the instruments. Progress with the construction of the telescope is continuing at our industrial partners in Europe and in Chile. With over 95% of the material budget spent, many components are in the manufacturing phases, including the many mirror segments and their mechanical supports for the main mirror, the remaining opto-mechanical components and parts of the dome and main structure. Significant work has been carried out on site, including the dome foundations on Cerro Armazones and a new technical facility as part of the Paranal observatory. As the telescope design evolves, a balance is sought between updating the interface information and maintaining the commitment to the numbers in the formal documentation. An informal, but controlled, exchange of information underpins the collaborative style that both ESO and the consortia wish to maintain while developing the most complex and costly instruments yet built for the most ambitious ground-based telescope ever. Workshops at ESO on the telescope and instrument operations concepts, on the alignment and verification of the instruments and on instrument software (pipeline, control and real-time control) have offered great opportunities for the exchange of the most up-to-date information between experts on the ESO and instrument teams.

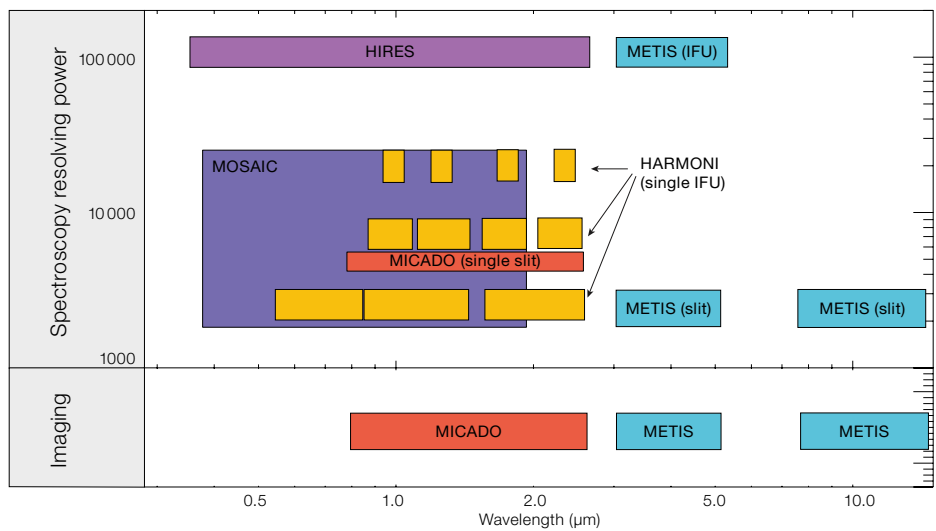

Figure 2. Parameter space for astronomical observations provided by the first-light and planned instruments on the Extremely Large Telescope.

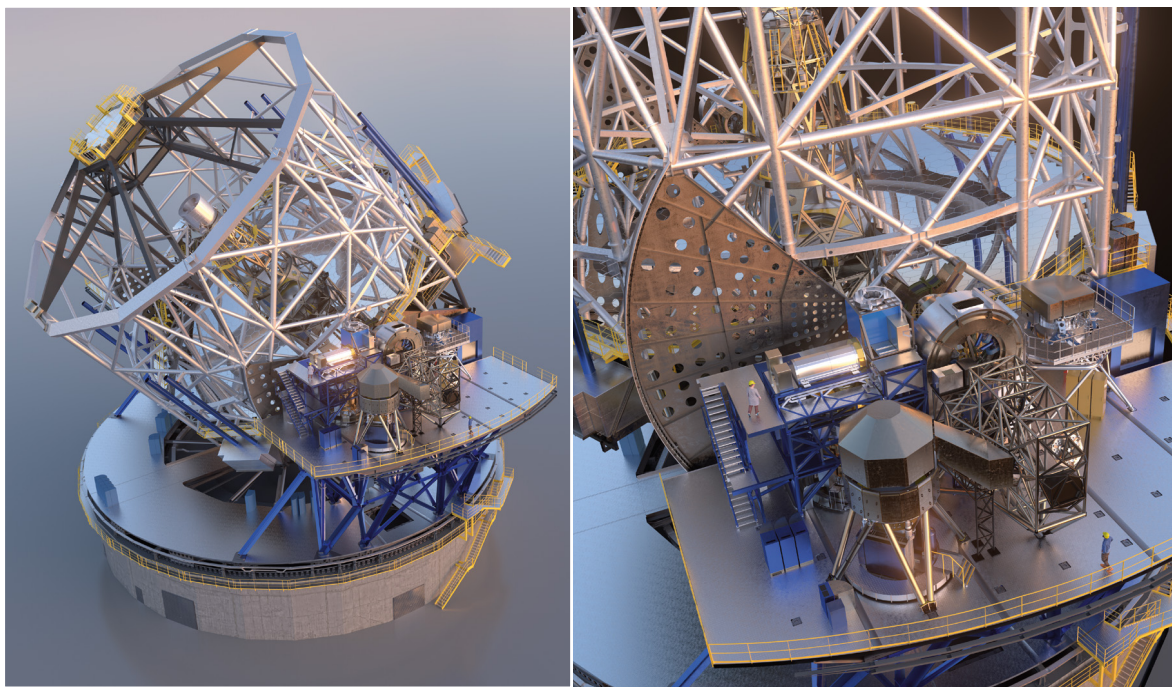

Figure 3. The instruments on the Nasmyth platform.

Part of the system-level activity at ESO is keeping an up-to-date model of the instruments on the Nasmyth platform, as the instruments themselves and the telescope main structure and prefocal station designs evolve. This model allows ESO and the instrument consortia to explore how to access various parts of the system during installation and maintenance, permits the dynamical modelling of the system under earthquake conditions and provides an all-important check that the instruments and other items on the Nasmyth platform do not occupy the same physical space or attachment points to the Nasmyth floor. The latest version of this layout is shown in Figure 3.

Looking towards the future operation of the ELT, a number of working groups[4] on specific topics have been set up. Membership of the working groups is open to anyone with an interest in contributing to the future scientific success of the ELT, whether from ESO, from the instrument consortia or from the community in general. The topics so far under discussion are: preparing for ELT observations (from observation preparation to execution); calibrations (including standard stars and astro-weather); calibration improvements and post-processing (including point spread function reconstruction and detector effect characterisation); and end-to-end modelling.

### Acknowledgements

Many more people than those listed as authors on this paper contribute to the development of the instruments for ESO's ELT. In particular, the importance of the work of the > 50 members of the follow-up team at ESO should not be underestimated. The authors would like to acknowledge the contribution of all those at ESO and in the community who are participating directly and indirectly in this exciting endeavour.

### Links

[1] ESO's Extremely Large Telescope (ELT): elt.eso.org
[2] The Thirty Meter Telescope (TMT): www.tmt.org
[3] The Giant Magellan Telescope (GMT): www.gmto.org
[4] ELT Working Groups: elt.eso.org/about/workinggroups/